# Paperstack – A Novel Lean-Interactive System for Documentation Sharing in Maritime Industries

Steinar Kristoffersen
Faculty of Computer Sciences
Ostfold University College
Moreforskning AS Molde; NO-6411 Molde
sk@hiof.no

The-Hien Dang-Ha*, Thien-Phuc Nguyen†
Faculty of Computer Sciences
Ostfold University College
Halden, Norway
hien.d.the@hiof.no*, phuc.nguyen@hiof.no†

*Abstract*—This paper defines a new domain for collaborative systems and technologies research: inter-organizational lean-interactive systems engineering. It departs from the problems pertaining to large-scale, technically demanding one-off construction projects, such as in our case studies, the building of specialized offshore service vessels. The requirements are unique, as are the explicit ambitions to be "lean", in other words, avoid waste and re-engineering by making sure that the process is customer-driven and rational. Emphasis is on efficiency rather than effectiveness. This set-up poses new demands to the collaborative systems and technologies, which we in this paper have addressed with the design of a new type of Documentation-emtric ERP, which we have called "Paperstack". The ambition is to support inter-organizational lean-interactive systems engineering in an integrated platform, and the next step for our research, naturally, is to put the systems into factual use. This paper summarizes the design ideas.

## I. INTRODUCTION

As complex networks of actors collaborate in design, engineering and construction, documentation sharing emerges as a crucial problem in the ship building industry. Each new ship requires thousands of documents (e.g. design drawings, engineering specifications, user manuals, or test certificates, etc...), which are collected from hundreds of different suppliers. Furthermore, the shipyards must supply all of these documents to ship owners before conducting the sea acceptance test as presented in contract. Research conducted in shipyards, with component suppliers as well as ship-owners showed us previously that current documentation-collecting process manifests many bottlenecks, which lead to late arrival, incompletion, and faulty format of documents [anonymous refs].

To understand how the problem is played out in real situations, we have taken part, as action researchers, in the team collecting documentation for two new ships (Recoded for the purposes of publication with No. 1 and 2) at ABC Ship Yard[1], where we assisted in organizing, copying, scanning, requesting and indexing the stipulated documents for about 370 working-hours in total. While directly involved in a real process, we revealed several major disadvantages of the current process, which is:

1) *Non-interactive*: No shared infrastructure between all the parties could be developed so far, because the maritime cluster is loosely coupled and all parties feel that they need to protect and control access to the flow of their information during projects. We sometimes had to scan pre-produced materials, duplicate the work by re-typing, or use deprecated material. These problems could have been handled with a shared infrastructure that supports access control, synchronization, and version management function, and in this paper we shall explore the pertaining design.

2) *Non-lean*: We did not know which documents were needed, when they should be requested, and how we should organize them systematically. Without full knowledge of which documents had been requested, sometimes all we could do was to passively wait for the suppliers to push their documentation our way and hope that they would provide all required documentation in time.

3) *Insufficiently conceptually aligned*, immersed in a *different working* "culture", and supported by *incompatible platforms* between ship owner, shipyard and suppliers: this amplifies the challenges on process.

The aim of this paper is to introduce a conceptual design of a novel lean-interactive system for documentation sharing that potentially solves these problems. By using SFI[2] classification system as a common information structure between different enterprises, we could implement a shared infrastructure that supports auto-synchronization, auto-request, access control, version management and communication between actors.

This paper intentionally ignores the technical issues related to how to implement the system such as data protection, network design, or synchronization through firewalls. The objective is to communicate the idea and its design.

## II. BACKGROUND

### A. Documentation Routines in Shipbuilding

Information in maritime industry is highly diverse, and many formats are being applied (e.g. PDF, Doc, CAD,...)

---

[1] Also recoded for the purpose of anonymous publication

[2] The abbreviation SFI reflects its origin in a Centre for Research-driven Innovation on Ship Building. The SFI Group System development was created and trialled for the first time in a pilot yard in 1972.

for a wide variety of purposes (e.g. specifications, drawings, plans,...), while it complexly flows across the value chain: the suppliers deliver components to different yards, while the yards in their turn have ships designed by various technical consulting companies [1]. Shipbuilding is a global business where work is traveling to the best and cheapest place around the world (e.g. those with labor-intensive such as steel work or assembly of the hull), while parts are collected and assembled close to operating place [2]. Therefore, shipbuilding's supply chain engages a large number of stakeholders and actors, who all have to supply up-to-date and relevant documentation for the ship before the completion stages.

Email or memory stick-based exchanges, which are currently widely-used solutions, will not provide an effective method to distribute documentation across systems, due to limited capacity, affordance of searching and real-time sharing, and version control let alone the needs of protecting, updating, or auto-requesting documentation. We have conducted research in shipyards, suppliers ( [1]), and ship owners ( [2]) to reveal the challenges behind the scene, which motivated us to propose the Paperstack solution for inter-supply chain documentation sharing.

*1) Challenges for suppliers:* Ship-component suppliers typically have thousands of customers, who have independent and potentially incompatible information infrastructures. Therefore, sharing documentation to all of these customers is challenging and costly, especially when suppliers want to protect and control access to their documentation as well. In the maritime industry, there is a loose coupling of companies, where partners today may be competitors tomorrow. However, there is still a lot of trust between different companies, which is maintained on the level of longitudinal and personal relations. Many suppliers do not maintain a DRM (digital rights management) scheme for the information flowing outside the enterprise network [1], which would enable them to build a shared infrastructure for partly-confidential documentation. In order to be able to deal with bigger project of an even more international nature, as the industry continues to go through globalization, we assert that automation and documentation control needs to become and integrated part of process governance, even in this industry.

The next challenge is that suppliers usually revise and update their documentation, which leads to the needs to do version management and implement a synchronization method to distribute the new version automatically and avoid circulating deprecated materials. Besides traditional approaches, which cannot sustain multi-versioned documents such as email or memory-stick exchanges, several companies provide external access to their own documentation management system, which due to its proprietary legacy cannot usually become a standard solution.

*2) Challenges in shipyards:* Besides the needs of maintaining the lose-coupling and managing different versions, the shipyards also encounter challenges in requesting, organizing, and forwarding documentation. Current documentation-collecting processes are non-lean, where suppliers are pushing documentation, for which the customer downstream (the shipyard) are not in control of specifying, rather than shipyards pulling it on demand. Instead, shipyards usually compile documentation as and when they get it, and if there is not, they request and re-request it until they get some. This property leads to haphazard arrival, data redundancy, and impossibility of collecting and organizing documentation in a systematic manner. Sometimes, shipyards do not know what and when to request, so they have to wait for suppliers and hope that they would provide all required documentation: "...we can never feel quite confident that it is all there" (shipyard planner).

The current documentation-collecting process is also non-interactive, since the contractual relations of suppliers-customers are implemented throughout a value chain that is highly specialized. When taking part in the documentation collecting team at ABC Ship Yard, we had to scan much of standard and pre-produced documentation (not unique for our ship), which suppliers could have sent to us easily (e.g., the user manual for a laundry machine), or work on documentation that had been taken care of by others already.

Furthermore, incompatible platforms and different categorization/index system between enterprises amplify the challenge. Each company deploys a particular documentation management system using different technology, which we could not ask them to change. Therefore, we need an international standard classification system, which could be used by any ship-related company, to generate a common code for the flow of information between different platforms.

*3) Challenges for ship-owners:* Before building a new ship, ship-owners have to work with designers and consultants to create a set of specifications and choose an appropriate shipyard. During the building process, these specifications will be refined and updated based on negotiations between shipyard and ship-owners. After that, a third-party agent will use these documents to supervise series of testing (e.g. harbor acceptance trails, operational acceptance trails, sea acceptance trails) to make sure the ship is built according to the specifications. Being updated frequently and shared between so many partners make these specifications hard to be maintained by just a single party. They need to be updated consistently and shared freely among all related party.

Moreover, current practices show that ships usually carry less documentation than they may end up needing. Documentation selection criteria are informal and implicit, especially for engineering manuals when they are not needed until some kind of repair or modification on the vessel. They mainly rely on consultants who having previous sailing experiences with a similar ship to distinguish between useful and useless documentation [2].

On board documentation is also not well organized. The owners sometimes do not know what they have, and may request it again from the yards or even make a new one by another consultant.

## B. The SFI Coding and Classification System

First released in 1972 by the Ship Research Institute of Norway, SFI (Skipsteknisk Forskningsinstitutt) is an international-standard classification system for maritime and offshore industry [3], [4]. There are currently more than 6000 SFI systems installed all over the world, using by many different companies from shipyards to offshore or consultancies companies [5]. The SFI system is a common ship breakdown system, which is applicable to all users with all types of ships, easy to understand, and capable of future expansion (described as the basic criteria for designing the SFI group system). The main purpose of SFI is to provide a common code for the flow of information within or between different enterprises in the maritime and offshore industry [3]. Ship-related companies could use SFI when dealing with information on ship specification, drawings, filing, maintenance, or many other purposes.

Designed as a function-oriented classification system, SFI uses a 3-digit decimal code to classify all functions involved in ship and rig operation. It divides ship/rig into 10 Main Groups from 0 to 9 (only Main Groups 1 to 8 are in use while 0 and 9 are used for uncovered components), each Main Group (1st digit) consists of 10 Groups (2nd digit), and each group is divided further into 10 Sub-Groups (3rd digit) [3], [4], [6]. For instance, if a part has SFI group number as 362, this indicates that it belongs to Main Group 3: "Equipment for cargo"; Group 36: "Freezing, refrigerating, and heating systems for cargo"; and Sub-Group 362: "Freezing and refrigerating systems for dry cargo" (Fig. 1).

In order to reach the component level, the SFI breaks the Sub-Groups further down by using 3-digit detail and material code. Detail code refers to components purchased directly to ship and acts in the sub system function, while material code relates to material purchased to stock. Code from 000 to 099 is used for detail code and 100 to 999 for material code. We put the detail/material code after its SFI group code to get the full 6-digit code. For instance, the SFI code 362.003 (3 digits of SFI Group code + 3 digits of detail code) refers to cooling compressor.

## III. SYSTEM DESIGNS

After analyzing all of these challenges carefully, we have designed the Paperstack system, which is an improved lean-interactive documentation system. We took that name as an indication of what we sometimes still see as the status quo of documentation work in many organizations, albeit one that the Paperstack system is intended to replace. The main objective of the system is to support so-called lean [7], [8] practices towards document completion (i.e., on-demand pulled rather than pushed, and only value adding operations and formats in documentation), based on the associations between information flow with value flow (e.g. design, engineering, and operations) of advanced maritime vessels. All documentation might then be shared freely among the partners who partake in a project, and would be pulled automatically whenever required, based on their project plans. All required materials are to be found in the right space, for the right place and at the right time. We did not expect to be able to standardize all flows of information within maritime cluster, since it is loosely coupled, highly specialized and many decisions are related to strategic concerns and business undertakings. However, all pre-produced and non-confidential documents should flow freely and logically throughout Paperstack to where they are needed. An indication of its success would be if one should not be able to find a documentation collector scanning the user manuals for a laundry machine, as what we did some months ago.

In the following sections, we are going to explain how we integrated the SFI classification system into the "electronic Paperstack" and how we modularized it into five functional modules to achieve the goals that we set out.

### A. Applying SFI in an "Electronic Paperstack"

In the Paperstack system, the SFI number is used as a common catalogue code and implemented as physical catalogues on a shared, global storage. In order to "stay within order", all the supply chain actors and project participants must maintain a SFI tree on their local disks to classify and store documentation, which might be synchronized into or from a global SFI tree (Fig. 1). A part (e.g. ship components, equipment, materials) is globally identified in the Paperstack by three attributes: The SFI number (location in SFI tree), part's name, and supplier's id (producer). To enable Paperstack to work in practice, almost all suppliers or vendors are required to use or integrate with the SFI system to classify and provide SFI codes for all their products, which is, fortunately already a common practice.

An element of the SFI tree is viewed as a folder, where all of its contained documents could be synchronized to different SFI structures when their identification attributes match. Concretely, when purchased a part, shipyards can add it into their local SFI trees using part's name, SFI code, and supplier id provided by supplier. After that, all documents of this part could be synchronized automatically when needed. In Paperstack, SFI trees are managed by a separate agent, which we have called the *PlaceFinder* module, while synchronization and requesting plan are handled by another agent, called the *FlowFinder*.

Sharing takes place at the level of a SFI identified location and downwards in the hierarchy, in other words, access should be granted collaborators to documentation of a build that was uploaded and indexed to a certain location that they share, and supported with familiar inheritance and scope rules to allow more partners to be added downwards, and names to be unique on that sub tree of the SFI representation of the build, rather than globally.

For prototype purposes, we suggested using *Rsync* onto linux disks, and encouraging a simple integration on the partner network side using a physical catalogue in the DMZ (de-militarized zone) onto which finished documents may be transferred, thus using a secure file transfer protocol based on ssh and very robust encryption.

### B. System's Modules

The overall Paperstack system is split into five following modules, or agents:

a) *iFinder*–connects existing documentation platform (including ideas, drawings, designs,...) to the vessel. A ship now can inherit all documentation from its ship type. Concretely, when creating a new ship that belongs to an existing ship type, a new SFI tree is also generated with all inherited documents in newest version.

b) *PartsFinder*–allows a shipyard to define and connect parts into ship's designs in order to generate parts list that needed to be documented. Shipyards now know exactly what document they need. During the engineering process, shipyards can revise, modify design and concretize parts until they are identifiable. In our system, a part is identifiable when it has SFI number, supplier id and part name.

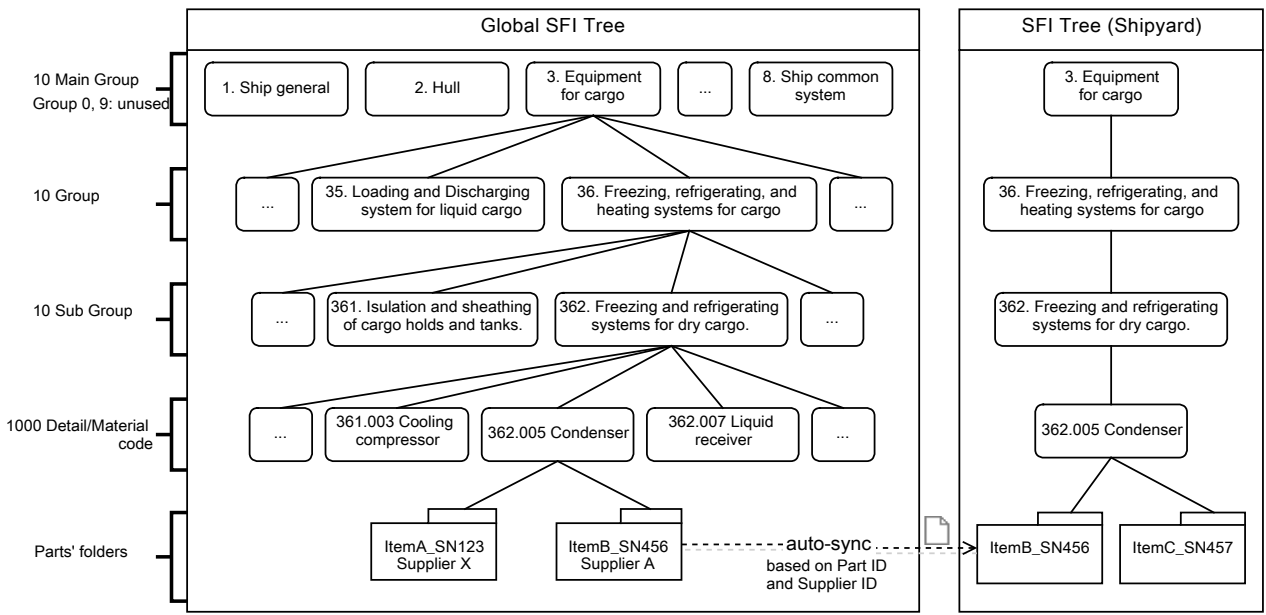

Figure 1: SFI tree structure and synchronization manner

c) *PlaceFinder*–manages SFI trees, which supports uploading, synchronization, access control, and version management on documentation. All parts defined by PartFinder are added into SFI tree of PlaceFinder and synchronized with the global SFI tree automatically. Suppliers can upload or update their documentation and grant right to their partners to have it synchronized into their local SFI tree.

d) *FlowFinder*–the most important module in the paperstack system, which takes responsibility of creating requesting plan, auto-synchronizing documentation, auto-requesting suppliers and auto-alerting shipyards. To accomplish these tasks, shipyards first have to input or import project plan from ERP into paperstack system. After that, they attach designs or parts into task that require them. This enables the system to analyst the flow of tasks and creates the suitable documentation-requesting plan. Concretely, the FlowFinder agent will request and synchronize documentation when shipyards need them for a particular task, and will alert users who are involved in this part of the process, when problems arise.

e) *iPartner*–set up a communication channel for actors involved in a particular document. In case synchronization could not be done automatically, shipyards can manually request document or negotiate about its right.

We provide a sequence diagram (XML 2.0) in Fig. 2 to visualize how the system should work in practice. To begin a new shipbuilding project, shipyards first find a suitable ship type and define a new ship inheriting all documents from it. After that, they iteratively use PartsFinder to concretize and attach parts to designs and SFI tree. They can add production plan (attached with designs and parts) into FlowFinder to get it generate the requesting plan. Based on that requesting plan, FlowFinder would ask the suppliers to upload required documents, grant right, or just synchronize it from global SFI tree if possible. It also alerts the shipyards when any problems or potential problems detected.

To make the system more intuitive and adaptive, we designed the main interface of Paperstack for suppliers as presented in Fig. 3. This interface supports suppliers when they:

- Search and browse their parts on SFI tree
- Upload, and manage documents.
- Check parts history and control access.
- Monitor events (e.g. requests, messages, alerts )
- Have conversation with shipyards.

*C. Information Flow Analytics*

Although the process of building an offshore vessel can be divided into stages, e.g. design, construction, building and commissioning; and a designated partner may be responsible for each stage, this is not the case in real projects. In a project, information does not flow from stage to stage following the conventional waterfall model. In fact, it forms a dynamic complex network, where every partner may need to exchange documentation with all others, and connections between them keep changing constantly based on changes in projects (e.g. stage changes, new negotiations, or changes in contracts). Indeed, even within a particular organization, information flow also gets complex when moving from one stage to another:

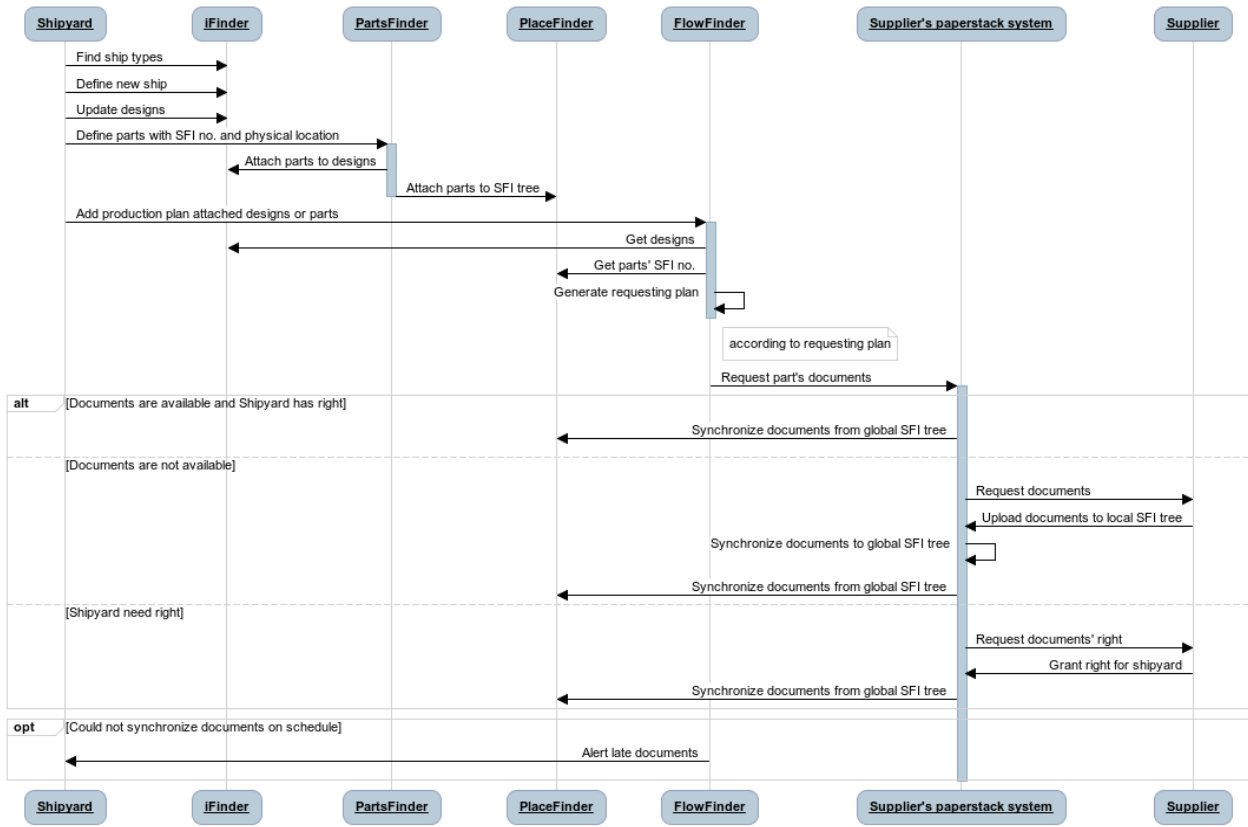

Figure 2: Paperstack sequence diagram

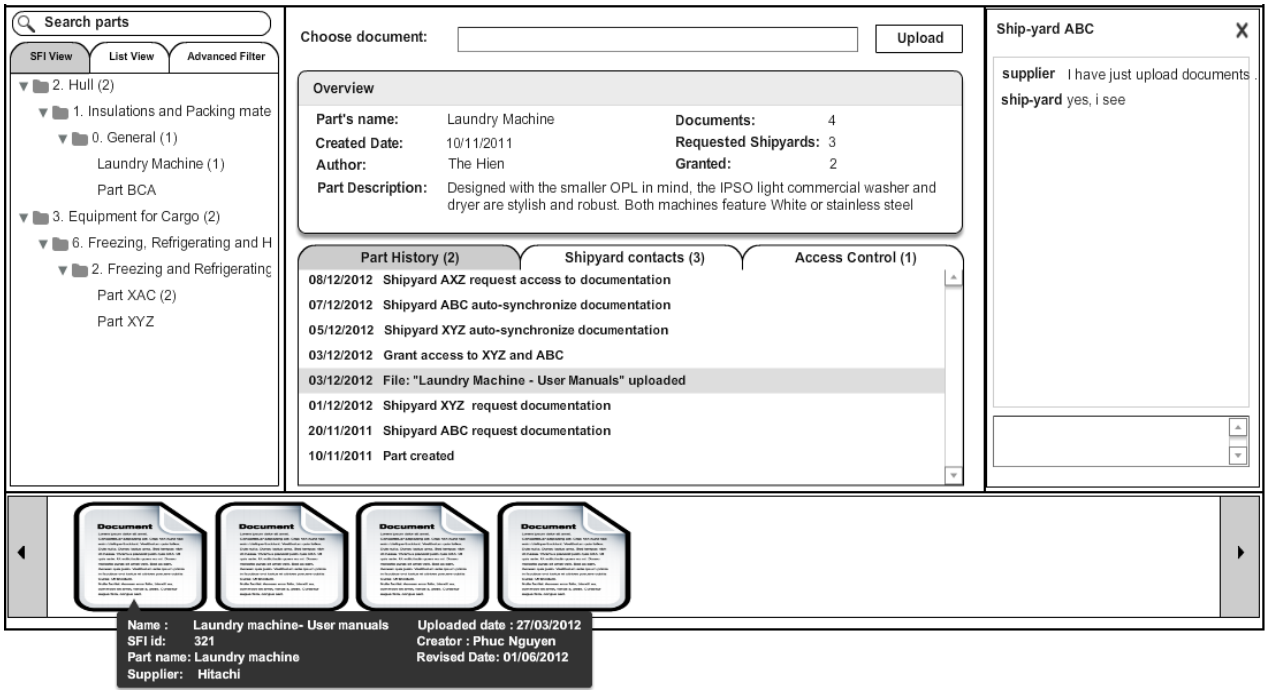

Figure 3: Mockup for Paperstack supplier view.

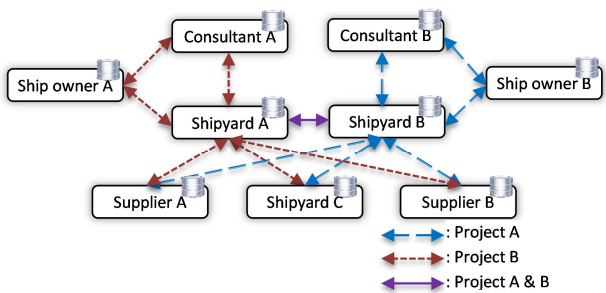

Figure 4: Simple maritime information network.

documentation is sometimes reproduced due to problems in information connection between different teams. Furthermore, each organization is also likely to partake in many different networks for different projects, and possibly take different roles. Information needs to be exchanged between organizations and across functional stages continuously. In Fig. 4, we depicted a snapshot of a simple information network for two projects A and B. Participants take different roles in different projects and have their own system to manage and share information with other partners.

We aim to improve the performance of this complex information network by using the Paperstack system as a digital information hub (Fig. 5). This system would enable a hub-like stakeholder behavior, by which integrated documentation flows make the latest versions of known documentation items ready exactly as and when they were needed. It standardizes the methods for exchange and storage maritime industry information. Information flow between projects, internal teams, and organizations now can run sufficiently freely, yet controllably to external partners through a single central Paperstack hub. It collects documentation from local storage, keeps it online, and distributes it dynamically and automatically according to the need of participants and changes of projects.

However, from a contractual perspective, there is a dissociation of the role of supplier and customer, the shipyard becomes a customer itself, taking the middle role of mediator between the suppliers who are able to provide documentation and the customer who have contractually staked a claim to it. Therefore, documentation needs to be polled from multiple sources, sometimes repeatedly. These polls may be of different types, for instances:

- Requesting the needed documentation: manual communication.
- Automated requests using the ERP (Enterprise Resource Planning): automatic communication.
- Contractual obligation: implicit communication
- Mediated communication, using for instance telephone or meetings, which would be polling as well: explicit communication

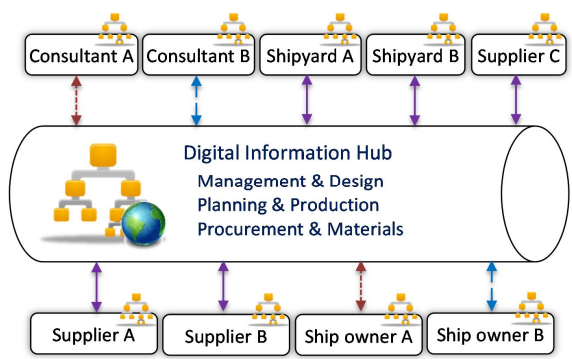

Figure 5: Digital information hub.

Thus we get the following polling matrix, which may be useful to structure the design:

TABLE I: Polling matrix

|  | Manual | Automatic |
|---|---|---|
| Implicit | Contract | Share |
| Explicit | Request | Workflow |

We wish to turn towards the automatic and implicit dimension and support interactive knowledge management through an open information space. However, a simple solution that documentation is replicated onto a shared server could not adequately address the challenges. We do not know necessarily from the beginning, the identity of each document, when we need it, and where we can get it. We believe that Paperstack may counter this, thanks to these modules:

- *PartFinder:* identifies the concrete element, job or mechanism that has to be documented for the delivery to be judged as complete. (What)
- *PlaceFinder:* keeps track of documentation belonging to a particular physical or logical space using the SFI classification scheme. (Where)
- *FlowFinder:* analyzes the building plan to indicate when we need a particular document. (When)
- *iPartner:* connects actors working together on a build. It supports explicit communication in case automatic-implicit document requests do not work.
- *iFinder:* helps shipyards find and inherit documentation for a new ship from the old similar ones.

## IV. CONCLUSION

This report describes the study and design implications, which came out of a project at ABC Ship Yard concerning the documentation completion phase of a new build two new ships in a series. The documentation process is itself documented through this report, including analytics about its problems in suppliers, shipyards and owners. However the main thrust have been towards designing a novel Paperstack system, which comprises designs as well as routines, which may speed up the process, as well as making it more robust.